# Single-atom level determination of 3-dimensional surface atomic structure via neural network-assisted atomic electron tomography

Juhyeok Lee, Chaehwa Jeong and Yongsoo Yang[*]

*Department of Physics, Korea Advanced Institute of Science and Technology (KAIST), Daejeon 34141, Korea*

## Introductory paragraph

Functional properties of nanomaterials strongly depend on their surface atomic structure, but they often become largely different from their bulk structure, exhibiting surface reconstructions and relaxations[1–9]. However, most of the surface characterization methods are either limited to 2-dimensional measurements or not reaching to true 3D atomic-scale resolution[5,6,8], and single-atom level determination of the 3D surface atomic structure for general 3D nanomaterials still remains elusive[10,11]. Here we show the measurement of 3D atomic structure of a Pt nanoparticle at 15 pm precision, aided by a deep learning-based missing data retrieval. The surface atomic structure was reliably measured, and we find that <100> and <111> facets contribute differently to the surface strain, resulting in anisotropic strain distribution as well as compressive support boundary effect. The capability of single-atom level surface characterization will not only deepen our understanding of the functional properties of nanomaterials but also open a new door for fine tailoring of their performance.

## Main text

Precise determination of 3D surface atomic structure at an individual atom level has been a main interest for broad scientific communities including physics, materials science, chemistry, and nanoscience. Due to lower coordination numbers, surface atoms often show substantial deviation from their bulk structure[1–4,7,9]. However, especially for metallic nanoparticles, the surface structure plays a crucial role in their catalytic activities, which nowadays have major technological importance in the synthesis of chemicals[12], abatement of air pollution[13], and fuel cell applications[14]. It is critically important to fully understand the surface atomic structure to fine-tune the catalytic properties for each application.

Atomic electron tomography (AET) has been recently developed as a powerful tool for individual atom level 3D structural imaging[15,16], actively being used for measuring atomic level defects[17,18]8/27/2020

7:06:00 PM, 3D strain[18,19], chemical order/disorder[17], and nucleation dynamics[20]. However, often due to geometrical limitations, only part of a full tomographic angular range is experimentally measurable (so-called "missing wedge" problem), which results in elongation and Fourier ringing artifacts along the direction of the missing information[21,22]. The missing wedge artifact strongly affects the accuracy of surface atomic structure, and it has been difficult to precisely measure the 3D surface atomic structure, being a main roadblock for precise determination of the 3D surface atomic structure[17].

On the parallel front, deep learning-based neural network approach has recently attracted great interest from electron microscopists[23–26]. It has already demonstrated successes in missing data retrieval[24,25] and super-resolution imaging[23,26,27]. In this work, we combined AET with deep learning-based neural network filter based on the *atomicity* principle. Using a Pt nanoparticle as a model system, we successfully retrieved the missing wedge information and achieved a robust reconstruction of the 3D surface atomic structure.

The main idea behind this deep learning-based approach is *atomicity* – the fact that all matter is composed of atoms. This means that the true atomic resolution electron tomogram should only contain sharp 3D atomic potentials convolved with the electron beam profile. Therefore, a deep neural network can be trained using simulated tomograms, which suffer from artifacts (due to missing wedge, insufficient projection data, various noises, etc) as inputs, and the ground truth 3D atomic volumes as targets. The trained DL network effectively works as a filter which removes the artifacts from tomograms. Figure 1 shows our deep learning (DL) filter architecture based on a 3D-unet[28] (Methods). Input training datasets are generated by simulating the electron tomography process using f.c.c. based random atomic models (Methods). Tomographic tilt series were obtained by linearly projecting the atomic potentials based on atomic scattering factors, and the broadenings due to electron beam profile and thermal vibration were also considered during the calculation (Methods). Only limited tilts angles (–65 to +65°) were used, and Poisson-Gaussian noises were added to simulate the experimental conditions. Three-dimensional tomograms were reconstructed from the tilt series using GENFIRE algorithm[29] (Methods). As shown in Fig. 1, undesirable artifacts along the vertical direction (missing wedge direction) were successfully simulated. These artifacts hinder defining the surface of nanoparticles and finding true atoms.

Using atom-tracing method, the 3D coordinates of individual atoms were determined from the simulated tomograms. Compared to the ground truth structure (Fig. 2a,f,k), the raw tomogram clearly suffers from artifacts resulting from the missing wedge and noise effect; the atomic intensities are blurred, elongated and connected to neighboring atoms, and several misidentified atoms can be found, especially near the surface (Fig. 2b,g,l). Applying the atomicity-based DL filter can successfully suppress the artifacts (Fig. 2c,h,m). The atomic intensities are well-localized, and most of the atoms are correctly retrieved. It can also be clearly seen from the FFT peaks that the missing wedge information was successfully restored (Supplementary Fig. 1) by the DL approach.

To mimic the true experimental conditions, including dynamic scattering, channeling, and lens aberrations, multislice-based PRISM simulations[30] were also performed to further test the DL filters (Methods). As expected, the raw tomograms from PRISM simulation show more artifacts near the surface compared to the linear counterpart (Fig. 2d,i,n). However, DL filter was still successful in reducing the artifacts even for PRISM-simulated tomograms, and the output volumes were as similar as the ones obtained from the linear simulations (Fig. 2e,j,o).

Since the traced atomic coordinates can be quantitatively compared to the ground truth atomic model, we performed a statistical analysis based on 1,000 test datasets, each generated from linear simulations and PRISM simulations. The raw tomogram tracing errors (percentage of incorrectly identified atoms; see Methods) of 6.8% (linear) and 7,6% (PRISM) were reduced to 0.4% and 0.6%, respectively (Supplementary Fig. 2a,b). Also, the before-filter root mean square deviation (RMSD) of 34.5 pm (linear) and 37.0 pm (PRISM) were substantially improved to 19.7 pm and 22.5 pm, respectively (Supplementary Fig. 2c,d).

To verify the robustness of our approach, the network was trained with two additional training datasets (one with f.c.c. based structure with different Gaussian width, and one with amorphous structure), and tested with the same test datasets used for the original DL filter. As can be seen in Supplementary Fig. 2, regardless of the base structure of the training data, all filters show consistent output, and proper f.c.c. ground truth structures of the test datasets were retrieved even by the filter trained with amorphous structure.

To evaluate the performance of our DL filter specifically in terms of surface structure, we calculated the tracing error and RMSD for the surface atoms before and after the filter. For the linear projection simulation case, the surface tracing error in Supplementary Fig. 2e decreased from 4.4% to 0.2%, and the surface RMSD reduced from 30.7 pm to 18.0 pm (Supplementary Fig. 2g). PRISM simulations also showed substantial improvement (tracing error 0.6%, and RMSD 21.3 pm after filter). These simulation results clearly demonstrate that the DL filter can successfully reduce the artifacts from insufficient data and noise, retaining improved precision especially for surface atoms. The filters trained with different training datasets also showed similar improvements (Supplementary Fig. 3).

We applied our DL filter to experimentally determine the 3D surface atomic structure of a Pt nanoparticle. The experiment was performed using an aberration-corrected scanning transmission electron microscope (STEM) operated in annular dark-field (ADF) mode (Methods). From a 4nm diameter Pt nanoparticle, a tilt series of 21 images were acquired with the tilt angles ranging from - 71.6° to +71.6° (Supplementary Fig. 4). After image post-processing, a 3D tomogram was reconstructed from the tilt series using GENFIRE algorithm[29] (Methods). Figure 3 shows the raw 3D reconstruction of the nanoparticle. Atom-tracing and classification procedure were applied to the volume, resulting in a 3D atomic model of 1,411 Pt atoms (Fig. 3c and Methods).

Severe artifacts due to the missing wedge problem and noise can be clearly seen in Fig. 3a and Supplementary Fig. 5, 6a-d. The reconstruction suffers from elongation and undesired intensity reduction near the surfaces, especially along the missing wedge direction. Although many of the atoms are correctly found at f.c.c. lattice sites, some of the atoms are clearly misidentified and surface atoms are not well defined (Fig. 3a,c and Supplementary Fig. 6). A 3D mask is often employed to define the surface of nanoparticles in this case. However, the mask depends on threshold parameters. Supplementary Fig. 6 shows the parameter dependence of the mask.

To resolve the precise atomic structure including the surface, the DL filter was applied to the raw tomogram. Figure 3d clearly shows that the atomic intensities in the filtered output are well isolated with expected Gaussian shape, showing drastic improvement compared to the raw reconstruction. Atom-tracing on the filtered volume resulted in 1,530 atoms; about 100 more atoms were successfully identified (Fig. 3c). Several missing atoms near the core region were restored by the DL filter (Fig. 3d and Supplementary Fig. 7). Averaged atom profile (Supplementary Fig. 8) clearly shows that the elongation artifact along the missing wedge direction is successfully resolved by the filter. Furthermore, the surface boundaries, especially along the missing wedge direction, are now clearly defined after the DL filtering, which allows unambiguous determination of 3D surface atomic structure without parameter-dependent masking process (Supplementary Fig. 6).

PRISM tomography simulation of the atomic model obtained from the filtered volume demonstrated 98.8% accuracy of atom identification and 15.0 pm precision of the atomic coordinates (Methods). We quantitatively analyzed the improvement made from the DL filtering (Supplementary Fig. 9). The atomic structures before and after the filtering showed the difference of 373 atoms (25.4%). About 60% of the difference comes from the surface (251 atoms). This mainly results from the unidentifiable surface atoms (due to missing wedge-induced intensity drop) being successfully traced after filtering (Supplementary Fig. 7). We further verified our approach by applying the two other filters trained by amorphous atomic models and f.c.c. atomic models with sharper Gaussian widths (Methods). The differences between the atomic models obtained from three different filters showed a higher consistency compared to the differences between the models before and after filtering (Supplementary Fig. 9), and the expected f.c.c. Fourier peak structures were well-retrieved even with the DL filter trained with amorphous atomic models (Supplementary Fig. 5).

Having accurate 3D atomic coordinates directly yields the 3D displacements and strain tensor. By comparing with an ideal f.c.c. lattice, the 3D displacements and strain tensor were calculated based on the traced atomic models (Methods). The out-of-plane atomic displacements in <100> and <111> facets (Fig. 4a-d) were -10.2 ± 62.9 pm and -3.3 ± 41.2 pm, respectively. However, part of the surface of the nanoparticle was making contact with the SiN membrane substrate. To understand the substrate effect, displacements of the facets making contact with the substrates were separately calculated, resulting in the out-of-plane displacements of -17.2 ± 86.5 pm for <100> facets and -21.7 ± 47.4 pm for <111> facets. For facets not in contact with the substrates, the average out-of-plane displacements of -6.4 ± 45.6 pm (compressive) and 5.3 ± 35.0 pm (tensile) were obtained for <100> and <111> facets, respectively. This behavior is consistent with the theoretical calculation result[3].

The strain map (Fig. 4) shows strong compressive strain along the x-direction and tensile strain along the y-direction. Interestingly, the strain along the z-direction shows both compressive (near the [001] facet) and tensile strain (near the [00$\bar{1}$] facet). The anisotropic strain behavior is likely to be related to the shape of the nanoparticle as well as the particle-substrate interface. Therefore, we conducted a shape analysis by assuming an ellipsoidal shape of the nanoparticle (Methods). The vertical direction (the shortest principal axis) of the nanoparticle is slightly tilted compared to the lab coordinate z-direction (i.e., electron beam direction), as indicated in Supplementary Fig. 10, and the lower right part of the nanoparticle makes contact with the SiN substrate.

The nanoparticle shape and facet analysis show that surfaces normal to x-direction are mainly affected by <100> facet atoms, and those normal to y-direction are mainly affected by <111> facets (Methods). This explains the opposite strain behavior along x- and y-directions. If we plot the strain map in the lab-coordinate (Supplementary Fig. 11), the z-directional strain $\varepsilon_{zz}$ exhibits a gradual change from tensile (the lower right part of the nanoparticle) to compressive (the upper left part). Although compressive strain is expected along z-direction due to the domination of <100> facets, strong tensile strain is observed at the particle-substrate interface. This result indicates that the nanoparticle surface structure is strongly dependent on the boundary condition, and the choice of support material can critically influence the structure, strain, and related catalytic behavior.

The strain maps were also calculated from the atomic structure obtained from the unfiltered tomogram and tomograms filtered with two other differently trained networks (using amorphous model, sharper Gaussian width model). Some differences can be found (especially near the surface) between the unfiltered case and filtered cases, but the strain maps from filtered tomograms are very consistent in overall, verifying the robustness of the DL filtering approach (Supplementary Fig. 11).

In conclusion, the 3D atomic structure of a nanoparticle was successfully determined at an individual atom level by neural network-assisted AET. Using a Pt nanoparticle as a model system, we demonstrated that the atomicity-based approach can reliably identify the surface atomic structure with a precision of 15 pm. The atomic displacement, strain and facet analysis revealed that the surface atomic structure and strain are related to not only the shape of the nanoparticle but also the particle-substrate interface. Combined with quantum mechanical calculations such as density functional theory, the capability of precise identification of surface atomic structure will serve as a powerful key to understand the surface/interface properties such as catalytic performance and oxidation effect.

## Data availability

Upon publication, all of our experimental data, image reconstruction, determined atomic structure will be posted on Materials Data Bank (https://www.materialsdatabank.org) open to public. All neural network-related codes and data analysis source codes will also be made available at http://mdail.kaist.ac.kr.

## Acknowledgement

We thank Chang Yun Son and Aloysius Soon for helpful discussions. This research was supported by the National Research Foundation of Korea (NRF) Grants funded by the Korean Government (MSIT) (No. 2019R1F1A1058236 and 2020R1C1C100623911). J.L and C.J. were also partially supported by the KAIST-funded Global Singularity Research Program (M3I3). The STEM experiment was conducted using a double Cs corrected Titan cubed G2 60-300 (FEI) equipment at KAIST Analysis Center for Research Advancement (KARA). Excellent support by Hyungbin Bae, Jinseok Choi, and the staff of KARA is gratefully acknowledged.

## Author Contributions

Y.Y. conceived the idea and directed the study. J.L. designed the neural network and performed simulational data analysis. Y.Y. designed and performed the experiment. C.J. and J.L conducted the experimental data analysis. J.L. and Y.Y. wrote the manuscript. All authors commented on the manuscript.

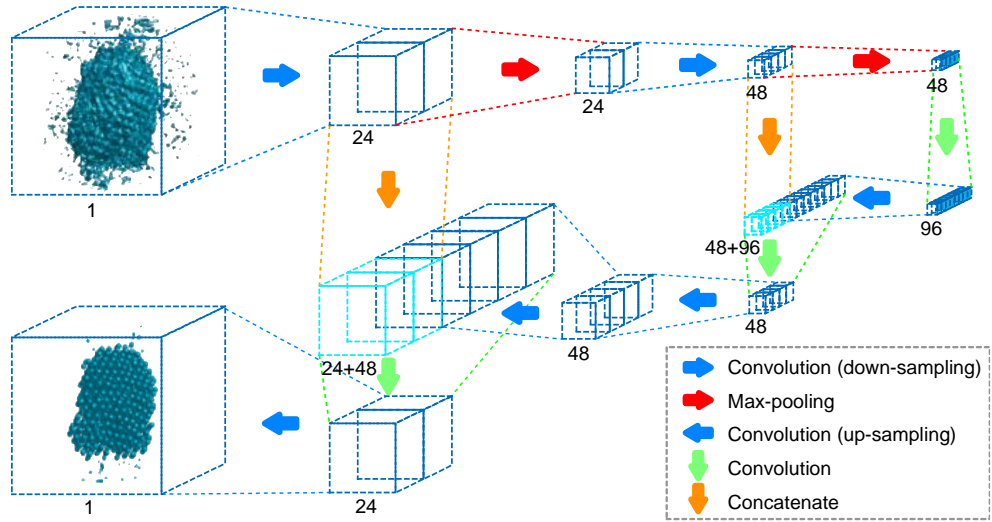

**Figure 1: The architecture of deep learning filter.** The deep learning filter follows a 3D-unet structure (Methods). The set of boxes represent feature maps. The number of channels is denoted below each feature map.

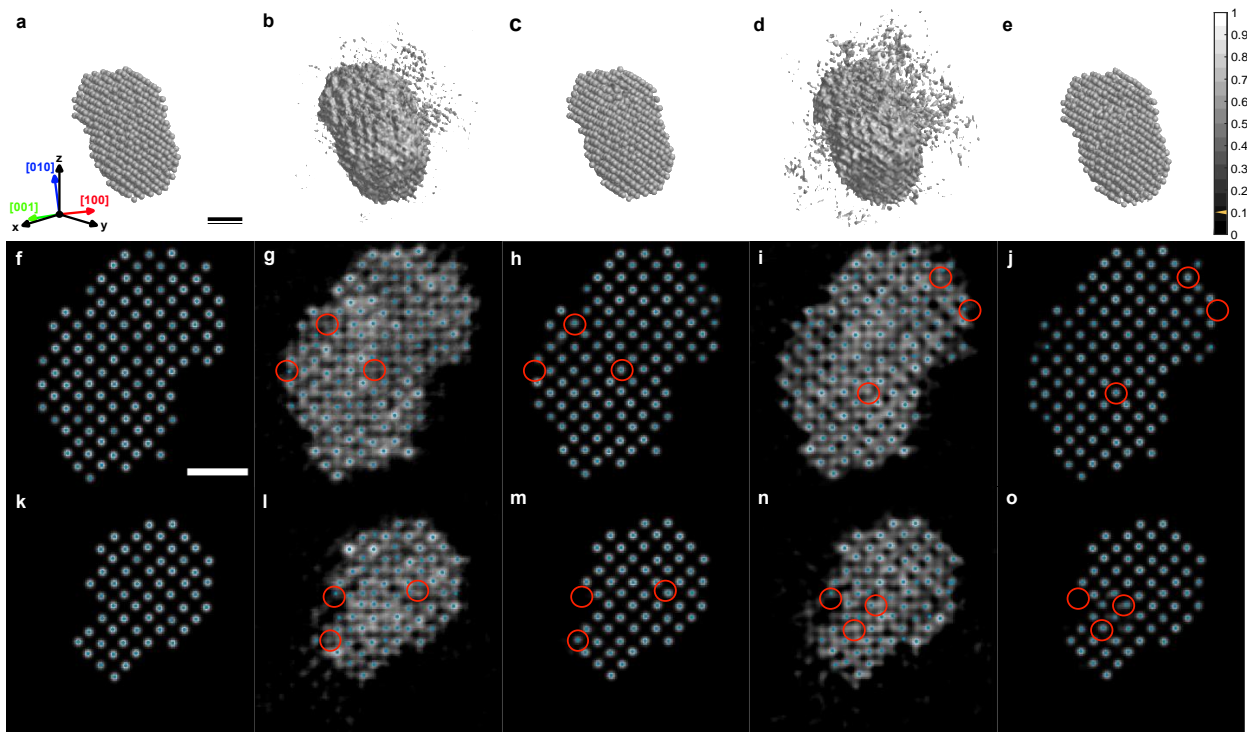

**Figure 2: Effect of the DL filter for simulated tomograms. a-e,** 3D iso-surfaces plotted with 10% iso-surface values (10 % of the highest intensity), representing ground truth (**a**), linear tomogram before (**b**) and after DL filter (**c**), PRISM tomogram before (**d**) and after filter (**e**). Note that the z-direction is the missing wedge direction. **f-o,** 2-Å thick slices of perpendicular to [001], obtained from the 3D tomograms near the center region (**f-j**) and near the surface (**k-o**): Ground truth (**f, k**), linear tomogram before (**g, l**) and after filtering (**h, m**), PRISM tomogram before (**j, n**) and after filtering (**i, o**). Grayscale background represents the reconstructed intensity, and blue dots represent the positions of traced atoms. Red circles denote misidentified atoms before DL filtering, which become correctly traced after the DL filter . Scale bars 1 nm.

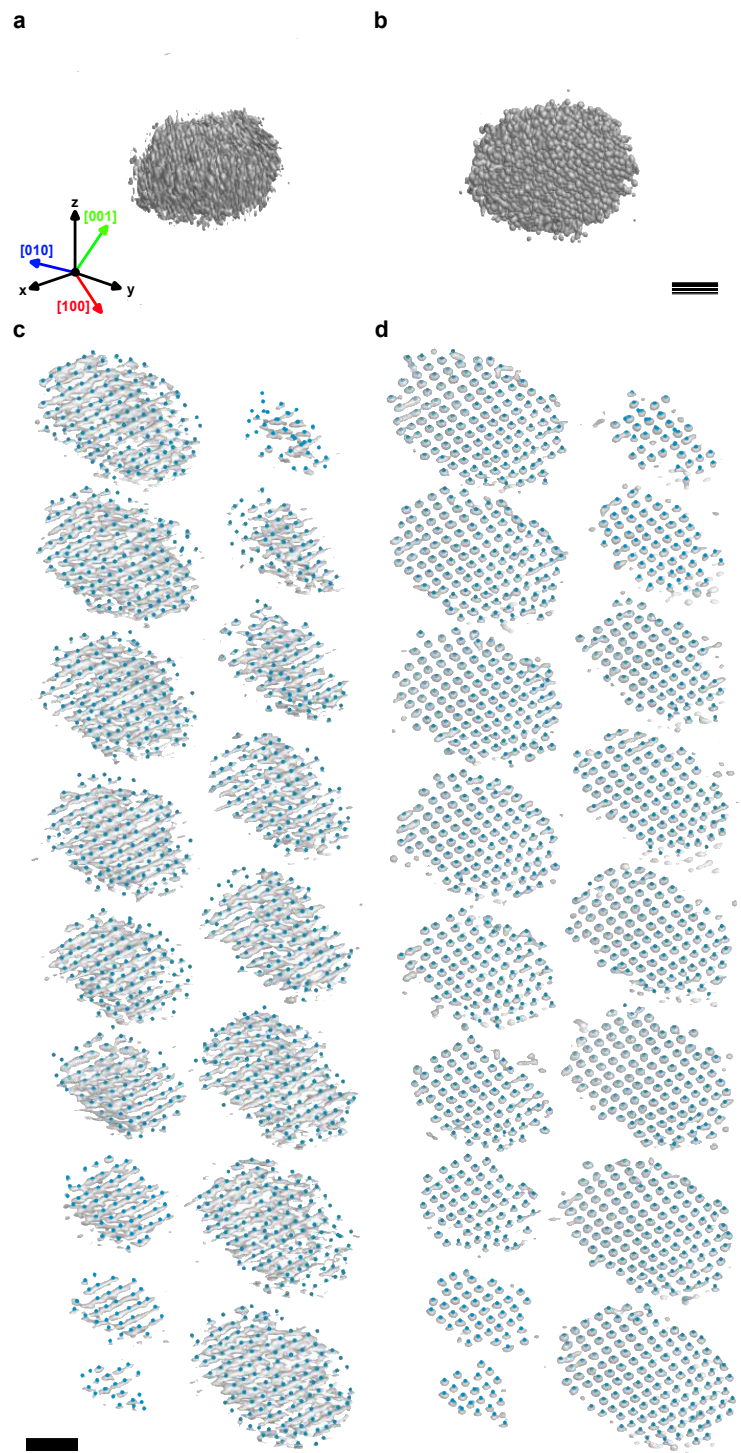

**Fig. 3: The 3D density maps of a Pt nanoparticle tomograms and traced atomic coordinates. a,b,** Iso-surfaces of 3D density are plotted with 40% and 15% iso-surface values from the maximum intensity before (**a**) and after (**b**) DL filtering, respectively. z-direction is the missing wedge direction. **c-d,** Raw reconstruction intensity and traced atom positions. Each slice represents an f.c.c. atomic layer, and the blue dots indicate the traced 3D atomic positions before (**c**) and after (**d**) DL filtering. The grayscale backgrounds of the atomic positions are iso-surfaces of 3D density with 40% (**c**) and 15% (**d**) iso-surface values from the maximum intensity. The sliced layers are perpendicular to [001] direction. Scale bars 1 nm.

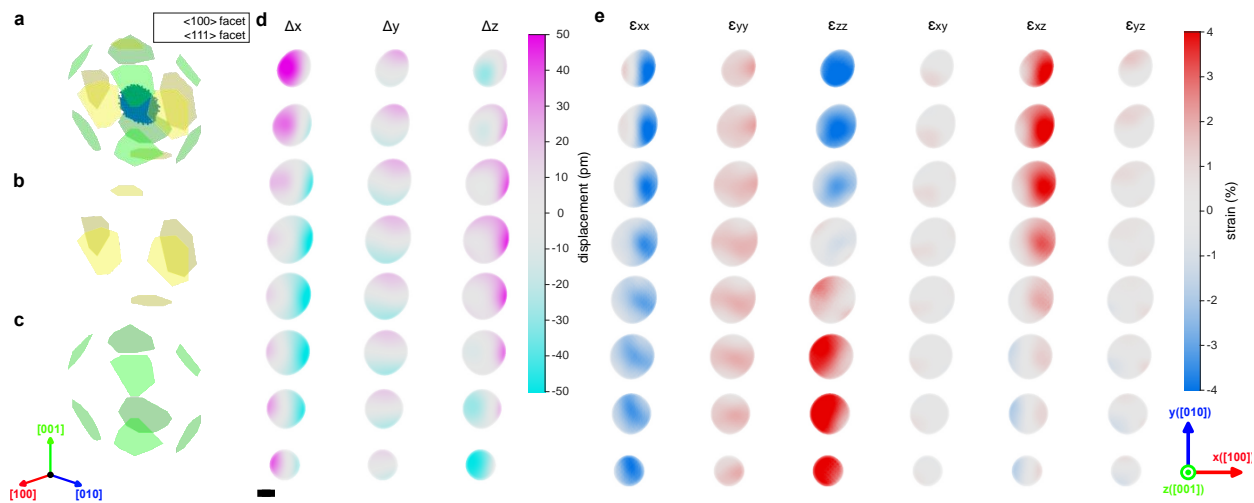

**Fig. 4: Facets, 3D atomic displacements and strain maps of the Pt nanoparticle. a-c**. Identified facet structure of the Pt nanoparticle, showing all facets (**a**), <100> facets (**b**), and <111> facets (**c**). **d**. The atomic displacements along the crystallographic axes **e**. The strain map representing six components ($\epsilon_{xx}$, $\epsilon_{yy}$, $\epsilon_{zz}$, $\epsilon_{xy}$, $\epsilon_{xz}$, $\epsilon_{yz}$) of the strain tensor. Scale bar 2 nm.